\documentclass[preprint,12pt]{elsarticle}
\usepackage{amsmath,amsfonts}
\usepackage{amsthm,amsmath,amssymb}
\usepackage{mathrsfs}
\usepackage{algorithmic}
\usepackage{array}
\usepackage{textcomp}
\usepackage{stfloats}
\usepackage{url}
\usepackage{verbatim}
\usepackage{graphicx}
\usepackage{balance}
\usepackage{xcolor}
\usepackage{epstopdf}
\usepackage{multirow}
\usepackage{booktabs}
\usepackage{latexsym}
\usepackage{algorithm}
\usepackage{hyperref}
\usepackage{marvosym}

\usepackage{amssymb}
\usepackage{subcaption}
\usepackage{nomencl}
\usepackage{multicol} 
\usepackage{tcolorbox}
\usepackage{times}
\usepackage{lineno}
\usepackage{color}

\journal{Applied Energy}

\begin{document}

	\begin{frontmatter}
		
		\title{Optimal Frequency Support from Virtual Power Plants: Minimal Reserve and Allocation}
		
		\affiliation[inst1]{organization={Department of Automation},
			addressline={Tsinghua University}, 
			city={Beijing},
			postcode={10084}, 
			country={China}}
		
		\affiliation[inst2]{organization={Laboratory for Information and Decision Systems (LIDS)},
			addressline={Massachusetts Institute of Technology}, 
			city={Boston},
			postcode={02139}, 
			country={United States}}
		
		\author[inst1]{Xiang Zhu}

		\author[inst2]{Guangchun Ruan}

		\author[inst1]{Hua Geng\corref{corresponding}}
		\ead{genghua@tsinghua.edu.cn}
		
		\cortext[corresponding]{Corresponding author.}

		\begin{abstract}
			
			This paper proposes a novel reserve-minimizing and allocation strategy for virtual power plants (VPPs) to deliver optimal frequency support. The proposed strategy enables VPPs, acting as aggregators for inverter-based resources (IBRs), to provide optimal frequency support economically. The proposed strategy captures time-varying active power injections, reducing the unnecessary redundancy compared to traditional fixed reserve schemes. Reserve requirements for the VPPs are determined based on system frequency response and safety constraints, ensuring efficient grid support. Furthermore, an energy-based allocation model decomposes power injections for each IBR, accounting for their specific limitations. Numerical experiments validate the feasibility of the proposed approach, highlighting significant financial gains for VPPs, especially as system inertia decreases due to higher renewable energy integration.
		\end{abstract}

		
	\end{frontmatter}

	\section{Introduction} \label{sec:intro}
	\subsection{Background and Literature Reviews}
	
	With the low-carbon transformation of power systems, an increasing number of renewable energy sources (RES), such as wind turbines and photovoltaics, are being integrated into the grid \cite{Back-lowinertia1,backgroud,Add-3}. Consequently, the inertia of future power systems will gradually decrease due to the growing integration of RES via power electronic interfaces, such as inverters~\cite{Back-lowinertia2, Back-lowinertia3,Add-4}. This reduction in system inertia may lead to deteriorated frequency stability if effective control measures, such as fast frequency regulation, are not implemented. To address this challenge, there is a general consensus on utilizing inverter-based resources (IBRs) to deliver ancillary services through grid-forming control, including inertia support and primary frequency response \cite{Back-FFR1,Back-FFR2,Back-FFR3}.
	
	In order to enhance the flexibility potential of IBRs during frequency regulation, virtual power plants (VPPs) have become the focus of current research~\cite{Add-5}. Specifically, a VPP can aggregate and coordinate large-scale IBRs to provide rapid frequency support services and enhance the real-time system balancing~\cite{Back-VPP1, ruan2024data}. In the study \cite{Back-VPP4}, a VPP framework was proposed including controllable heat pump water heaters and batteries to provide frequency support in coordination with other generation units. The work in \cite{Back-VPP3} developed a self-motivated frequency regulation mechanism considering the privacy preservation policy for VPP operators. Besides, due to its quick and flexible regulation capabilities, the VPP has demonstrated significant economic potential, particularly within real-time electricity markets that operate on a timescale of seconds~\cite{ Back-VPP5,Add-6}.
	
	Recent studies have emphasized the role of VPPs in improving real-time electricity market performance through enhanced coordination, uncertainty management, and flexible response mechanisms. For example, studies such as~\cite{Market-3, Market-4} have analyzed the interactions between day-ahead and real-time market behaviors of VPPs, accounting for the uncertainties inherent in such markets. These uncertainties were addressed using robust optimization and stochastic optimization approaches, thereby ensuring more reliable participation of VPPs in energy trading. Furthermore, the work in~\cite{Market-2} proposed a real-time cooperation strategy that effectively coordinates wind power generation and battery storage within a VPP, with the flexible response capability of battery storage allowing the VPP to accurately track regulation signals in real time. Additionally, in~\cite{Back-VPP2}, the real-time market participation of a VPP including RES, storage batteries, and flexible loads was studied using the interval method to handle the available uncertainties. 
	
	Building upon the comprehensive discussions so far, a critical gap persists in the accurate determination of the reserve capacities for frequency regulation on the timescale of seconds, especially in the context of VPPs participating in real-time markets. In \cite{COM1}, appropriate power settings and reserve capacities for IBRs were determined with a five-minute step length to provide secure and cost-effective inertia support. In \cite{COM2}, a frequency-constrained stochastic optimization model was proposed to obtain optimal regulation reserves with a fifteen-minute step length, along with power dispatch strategies for IBRs. In \cite{COM3}, regulation reserves were evaluated with an hourly step length under different de-loading schemes to determine the optimal allocation of generation units. In summary, existing studies predominantly focused on longer timescales, such as minutes or hours, which did not adequately capture the rapid dynamics needed for precise frequency regulation, thereby hampering the effective integration of VPPs into real-time electricity markets. 
	
\begin{figure}
		\centering
		\includegraphics[width=0.8\linewidth]{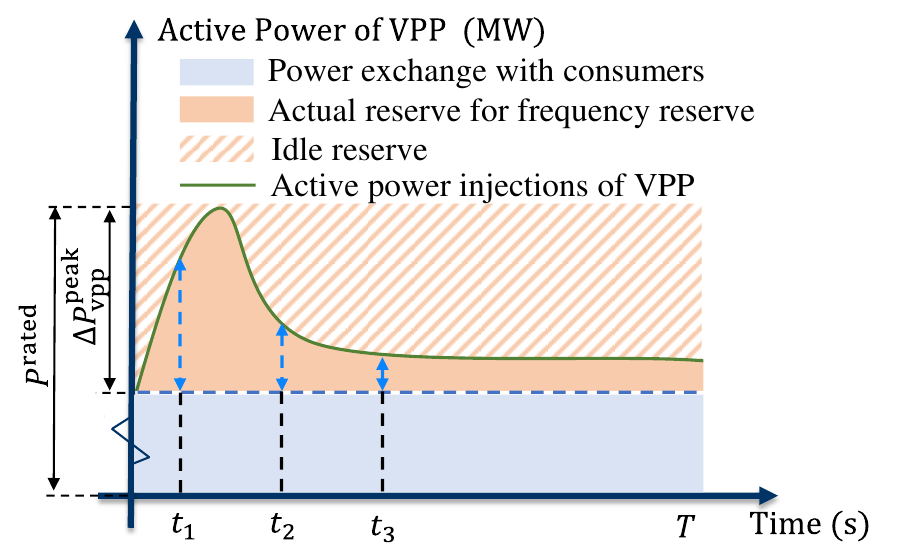}
		\caption{Fixed frequency regulation reserve ignores the dynamics on the timescale of seconds of VPP-level active power injections. The fixed reserve is determined by the peak value of active power injections, which contains the actual reserve and the idle one.} 
		\label{fig.framwork}
	\end{figure}

	Specifically, a VPP with fast-responding IBRs should adapt to other frequency response units in the power system, which may have inherently slower response dynamics, synchronous generators (SGs) for example. This mismatch in response speeds can lead to significant overshoots in active power injections from IBRs. Ignoring these rapid dynamics in existing reserve determination strategies that operate on timescales of minutes or longer \cite{COM1, COM2, COM3} results in a substantial portion of frequency reserves remaining idle, as illustrated in Fig.~\ref{fig.framwork}. This idle reserve portion highlights that traditional reserve strategies are overly conservative, failing to leverage the full potential of fast-responding IBRs and thus adversely affecting the financial opportunities for VPPs in the real-time electricity market.
	
	\subsection{Paper Contributions}
	
	To address the aforementioned research gaps, this paper proposes a reserve-minimizing strategy for VPPs to provide effective frequency support while enhancing real-time electricity market participation. The proposed approach determines the VPP-level minimal regulation reserves on the timescale of seconds that fully adapt to grid conditions and subsequently allocates these reserves to heterogeneous IBRs through optimization. By explicitly considering the second-level dynamics of active power injection, the strategy ensures the frequency safety of the system and significantly improves reserve utilization. 
	
	The major contributions are summarized as follows:
	
	\begin{itemize}
		\item[$\bullet$] {A reserve-minimizing strategy is proposed to align VPP reserves with actual power injections that are determined by virtual inertia and damping. By minimizing idle reserves, this approach enhances the flexibility of the VPP and facilitates better participation in real-time electricity markets.}
		\item[$\bullet$] {An adaptive feasible region is developed to determine the minimal frequency reserves based on the closed-form derivation of the power injections with second-scale fluctuations. The critical system constraints consider not only frequency safety metrics but also decay rate constraints.}
		\item[$\bullet$] {An energy-based optimal allocation model is formulated to allocate the determined reserves among heterogeneous IBRs. This model fully accounts for the capacity constraints and financial preferences of IBRs to achieve optimal economic benefits of the VPP.}
	\end{itemize} 
	
	\begin{figure}
		\centering
		\includegraphics[width=0.8\linewidth]{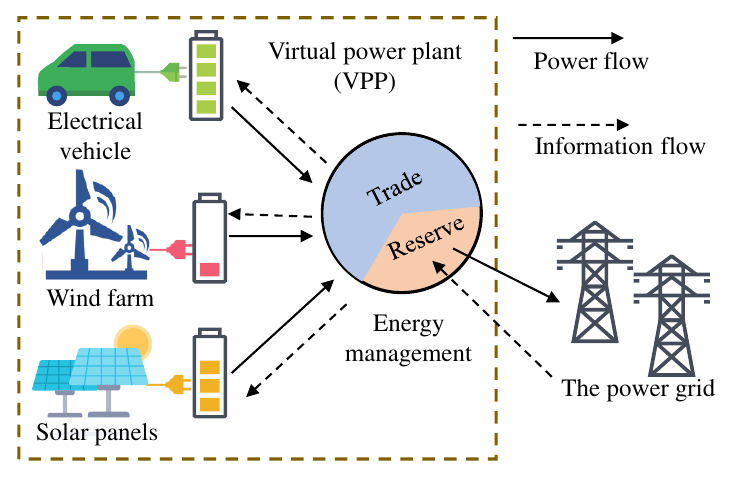}
		\caption{A typical VPP system hosts a group of IBRs, including electrical vehicles, the wind farm and solar panels with battery storage systems. The energy managed by VPP is used for trade (financial gains) or reserve (frequency support).} 
		\label{fig.compVPP}
	\end{figure}
	
	The rest of this paper is organized as follows. Section~\ref{sec:problem} explains the problem formulation. Section~\ref{sec:modeling} presents the modeling of the VPP-integrated system. Section~\ref{sec:strategy} derives the minimal reserves for the frequency support of the VPP. Section~\ref{sec:allocation} proposes an optimal allocation model for IBRs within the VPP. Case studies are presented in section~\ref{sec:cases}. Finally, section~\ref{sec:conclusion} concludes the paper.
	
	\section{Problem Formulation} \label{sec:problem}
	
	As shown in Fig.~\ref{fig.compVPP}, the VPP is allowed to participate in the real-time electricity market for trade and also support the grid frequency. Determining the frequency reserve is a crucial preparation step for the power grid, and the common option is the fixed reserve strategy that ignores the overshoot dynamics on a timescale of seconds. The shaded portion of capacity in Fig.~\ref{fig.framwork} is reserved but eventually idle. It should be pointed out that the reserve market is not considered in this paper, thus maintaining idle reserve will deteriorate the benefit of VPP participating in the real-time electricity market.
	
	We consider a VPP hosting a group of IBRs (see Fig.~\ref{fig.compVPP}) that can respond to frequency regulation signals. {Potential IBRs include electric vehicles, wind farms, and solar panels with battery storage systems respectively~\cite{Add-1, Add-2}. The uncertainty from renewable resources can be suppressed to a certain extent because battery systems can be seen as stable sources of energy on the timescale of seconds. The impacts of uncertainty and possible model errors will be discussed in Section~\ref{sec:allocation}.} The operational cost of IBRs varies widely, which motivates us to focus on the optimal allocation strategies among these heterogeneous IBRs for frequency reserve.
	
	Based on this foundation, this paper explores methods to enable the VPP to provide frequency support with minimal reserves through optimized reserve decision-making and allocation. To address this challenge, it is crucial to derive the regulation requirements that not only meet essential constraints but are also practical for allocation optimization. To this end, we conduct a detailed analysis of the demands of frequency regulation, quantifying the requirements for the VPP in a parametric and systematic way.
	
\begin{figure}
		\centering
		\includegraphics[width=0.7\linewidth]{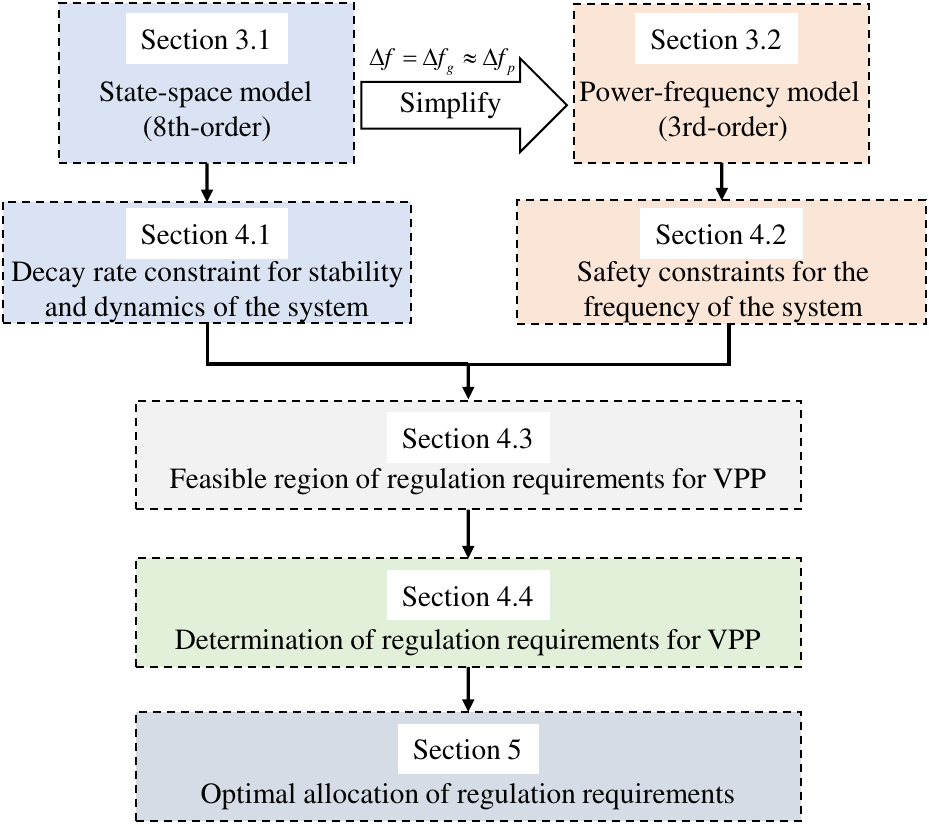}
		\caption{{Overview of the proposed method to derive and allocate the regulation requirements.}} 
		\label{fig.overview}
	\end{figure}

	Specifically, a system model incorporating VPP interactions is developed to analyze the frequency deviation mechanism and its associated regulation schemes. Based on this analysis, the frequency regulation requirements are derived, capturing the significant second-scale fluctuations in power injections from the VPP. These requirements are then quantified as reserves and optimally allocated to various IBRs. The framework of this study is illustrated in Fig. \ref{fig.overview}.

	\section{System Model with VPP Interactions} \label{sec:modeling}
	
	\subsection{State-Space Model for VPP-Integrated System}
	From an implementation perspective, the system can be actually divided into two parts, i.e., the VPP and the power grid and their connection point is called the point of common coupling~(PCC). The frequency response of the power grid, including SGs and loads, is modeled in the mode of the center of inertia~(COI). The VPP, coordinating IBRs with frequency regulation capacity, is equivalent to an aggregated power plant realizing synchronization with the grid through phase-lock-loop~(PLL) \cite{FFR1,FFR5}. The aggregation parameters of the VPP include virtual inertia $H_\text{VPP}$ and damping $D_\text{VPP}$ which are variables to be determined and other equivalent device-level control parameters obtained by measuring.  
	
	The frequency deviation of the grid, represented by $\Delta f_{g}$, is generated by the swing equation as in (\ref{fre_grid}). 
	\begin{subequations}
		\setlength{\abovedisplayskip}{3pt}
		\setlength{\belowdisplayskip}{3pt}
		\begin{align}
			& \Delta P_{e} + \Delta P_\text{PFR} = 2H_{0}\frac{{{\rm d}{f_g}}}{{{\rm d}t}}+D_{0}\Delta f_{g}\\
			& \Delta f_{g} = f_{g}-f_{0}	
		\end{align}
		\label{fre_grid}%
	\end{subequations}
	where $\Delta P_{e}$ is the power disturbance which is determined by (\ref{td-dis}), $\Delta P_\text{PFR}$ is the power injection from SGs in the grid side during the primary frequency response (PFR), which is determined by (\ref{td-SG}), $H_{0}$ and $D_{0}$ are the equivalent inertia and damping of the grid.
	
	\begin{equation}
		\setlength{\abovedisplayskip}{3pt}
		\setlength{\belowdisplayskip}{3pt}
		\Delta P_{e} = P^{G}+P^{V}-P^{L}
		\label{td-dis}
	\end{equation}
	where $P^{G}$ is the power output of SGs, $P^{V}$ is the power output of the VPP and $P^{L}$ is the power demand of loads.
	\begin{equation}
		\setlength{\abovedisplayskip}{3pt}
		\setlength{\belowdisplayskip}{3pt}	
		T^\text{SG}\frac{{\rm d}\Delta P_\text{PFR}}{{\rm d}t}+\Delta P_\text{PFR} = R(\Delta f_{g} + f_\text{DB2}), t\ge t_\text{DB2}
		\label{td-SG}
	\end{equation}
	where $T^\text{SG}$ represents the governor delay in time, $R$ is the equivalent droop coefficient of PFR from SGs, $f_\text{DB2}$ is the width of the dead band of SGs and $t_\text{DB2}$ is the time point that SGs start to provide PFR services.
	
	The frequency deviation of the VPP ($\Delta f_{p}$) is regulated by PLL through voltage measuring at PCC as in (\ref{fre_pll}). The device-level control strategies are analyzed and modeled based on the Park's Transformation \cite{Q-current}.
	\begin{equation}
		\setlength{\abovedisplayskip}{3pt}
		\setlength{\belowdisplayskip}{3pt}
		\frac{{{\rm d}{f_p}}}{{{\rm d}t}} = K_{p}^I{E_q} + K_{p}^PM({f_g} - {f_p})	
		\label{fre_pll}
	\end{equation}
	where $E_{q}$ is the q-axis voltage of PCC point, $K_{p}^{I}$ and $K_{p}^{P}$ are the coefficients of PI regulator and $M = 2\pi f_{0}$.
	
	Based on the measured frequency $f_{p}$, the current reference value can be determined as in (\ref{fre_vpp}) through an additional frequency control loop.
	\begin{equation}
		\setlength{\abovedisplayskip}{3pt}
		\setlength{\belowdisplayskip}{3pt}
		i_d^{ref} = i_{0}^V - {D_\text{VPP}}(\Delta f_{p}+ {f_\text{DB1}}) - 2{H_\text{VPP}}\frac{{{\rm d}{f_p}}}{{{\rm d}t}}
		\label{fre_vpp}
	\end{equation}
	where $i_d^{ref}$ and $i_{0}^V$ are the actual and setting current reference values of the d-axis, $\Delta f_{p}$ is the deviation of frequency and $\Delta f_{p} = f_{p}-f_{0}$. The reference value of the q-axis current is set as zero ($i_q^{ref} = 0$) to make sure there is no reactive current being injected into the grid to support the voltage under non-fault conditions \cite{Q-current}.
	
	\begin{figure}
		\centering
		\includegraphics[width=0.7\linewidth]{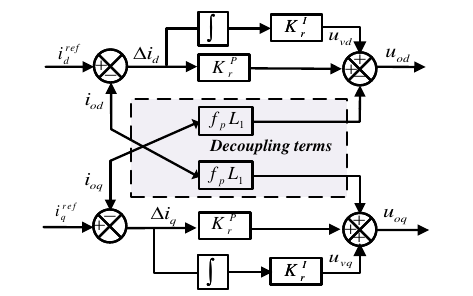}
		\caption{The current inner loop control diagram.} 
		\label{fig.currentloop}
	\end{figure}
	
	Besides, as in Fig. \ref{fig.currentloop}, the other states in the current inner loop are modeled as in (\ref{curren-1}).  
	\begin{subequations}
		\setlength{\abovedisplayskip}{3pt}
		\setlength{\belowdisplayskip}{3pt}
		\begin{align}
			& {u_{od}} = {u_{vd}} + K_r^P(i_d^{ref} - {i_{od}}) - {i_{oq}}{f_p}{L_1} + {E_d}\\
			& {u_{oq}} = {u_{vq}} + K_r^P(i_q^{ref} - {i_{oq}}) + {i_{od}}{f_p}{L_1} + {E_q}
		\end{align}
		\label{curren-1}%
	\end{subequations}
	where $u_{vd}$ and $u_{vq}$ are the outputs of current inner loop integrator and their dynamic represents as ${\rm d}u_{vd}/{\rm d}t = K_{r}^I({i_{d}^{ref}} - {i_{od}})$, ${\rm d} u_{vq}/{\rm d}t = K_{r}^I({i_{q}^{ref}} - {i_{oq}})$, $u_{od}$ and $u_{oq}$ are the outputs of current inner loop~\cite{Sun_2018}.
	
	Considering the coupling relationship between voltage and current, the state dynamic equations are as follows in (\ref{curren-2}).
	\begin{subequations}
		\setlength{\abovedisplayskip}{3pt}
		\setlength{\belowdisplayskip}{3pt}
		\begin{align}
			& {L_1}\frac{{{\rm d}{i_{od}}}}{{{\rm d}t}} = {L_1}{f_p}{i_{oq}} + {u_{od}} - {E_d} - {R_1}{i_{od}}\\
			& {L_1}\frac{{{\rm d}{i_{oq}}}}{{{\rm d}t}} =  - {L_1}{f_p}{i_{od}} + {u_{oq}} - {E_q} - {R_1}{i_{oq}}
		\end{align}
		\label{curren-2}%
	\end{subequations}	
	where $L_{1}$ and $R_{1}$ are the filter impedance and resistor respectively. Based on (\ref{fre_grid})-(\ref{curren-2}), the state-space model of the VPP-integrated system is established as in (\ref{eq.state-space}). In the per unit system, the active power of the VPP ($P^{V}$) is equivalent to the d-axis current $i_{od}$ due to $i_{oq} = 0$ and $u_{oq} = 1$ (p.u.). 
	\begin{equation}
		\setlength{\abovedisplayskip}{3pt}
		\setlength{\belowdisplayskip}{3pt}
		\bf{\dot x = Ax + Bu}
		\label{eq.state-space}
	\end{equation}
	where $\bf{x}$ is the vector of states of the eighth-order state-space model (shown in (\ref{state})) and the matrix $\bf{A}$ and $\bf{Bu}$ are listed in the Appendix.
	\begin{equation}
		\setlength{\abovedisplayskip}{3pt}
		\setlength{\belowdisplayskip}{3pt}
		{\bf{x}} = {\left( f_{p}, E_{q}, f_{g}, u_{vd}, u_{vq}, i_{od}, i_{oq}, \Delta P_\text{PFR} \right)}^\top
		%\mathbf{A}^\top
		\label{state}
	\end{equation}
	
	\subsection{Reduced-Order Model for System Frequency Response}
	
	\begin{figure}
		\centering
		\includegraphics[width=0.8\linewidth]{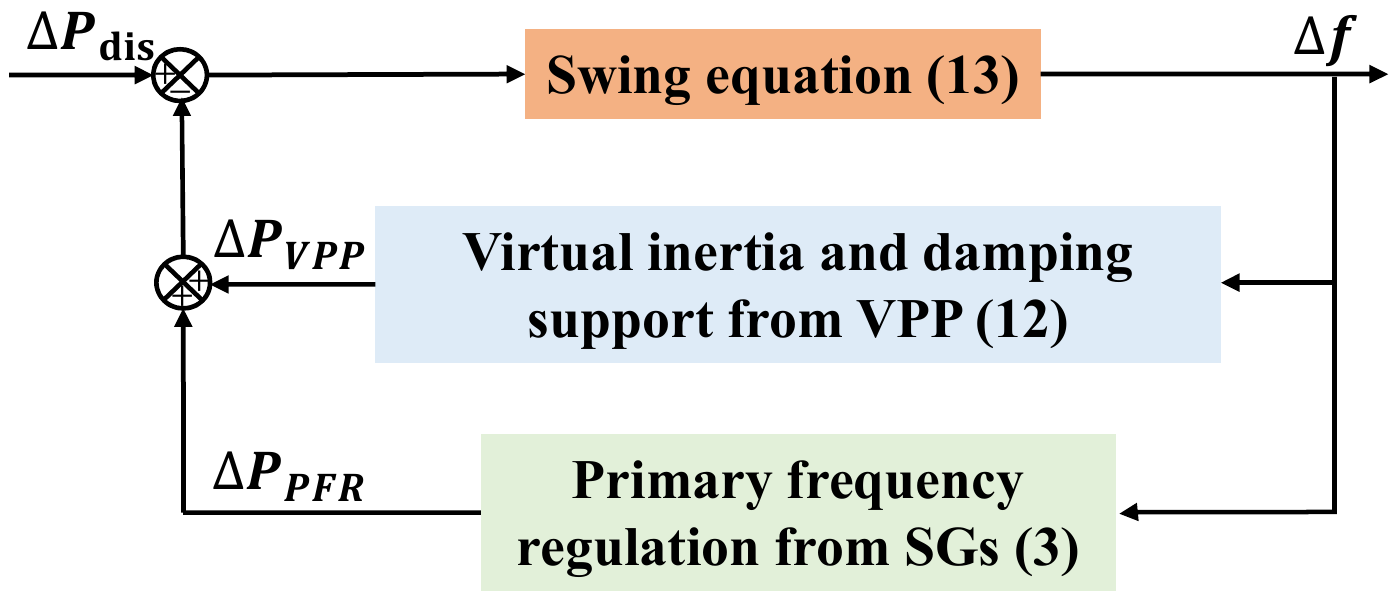}
		\caption{The block diagram illustrates the system frequency response. The forward path represents the relationship between power disturbance and frequency deviation of the grid. While the green part and blue part describe the frequency responses of SGs and VPP, respectively.} 
		\label{fig.block}
	\end{figure}
	
	To avoid solving the complex high-order equations derived in (\ref{eq.state-space}) directly, we carry out a model reduction to obtain the system frequency response. The main idea is to extract the relationship between the active power disturbance $\Delta P_\text{dis}$ and the system frequency deviation $\Delta f_{g}$. 
	
	Because of the fast regulated speed, the transient of the VPP tracking the grid frequency is short enough to be neglected~\cite{PLL2}. Therefore, it can be considered that the two are approximately homologous as in (\ref{approxi}).
	\begin{equation}
		\setlength{\abovedisplayskip}{3pt}
		\setlength{\belowdisplayskip}{3pt}
		\Delta f = \Delta f_{g} \approx \Delta f_{p}
		\label{approxi}
	\end{equation}
	
	Without loss of generality, we consider the frequency drop and derive the system frequency deviation~($\Delta f$) in (\ref{swing}) based on the swing equation.
	\begin{equation}
		\setlength{\abovedisplayskip}{3pt}
		\setlength{\belowdisplayskip}{3pt}
		\Delta P_\text{dis} = 2H_{0}\frac{\text{d}\Delta f}{\text{d}t}+D_{0}\Delta f
		\label{swing}
	\end{equation}
	
	It should be pointed out that the equation (\ref{swing}) describes an open-loop relationship between the power disturbance and frequency deviation without regulations from VPP and SGs.	To enhance the frequency safety, effective frequency modulations are considered, including the PFR of SGs (\ref{td-SG}), virtual inertia and virtual damping provided by the VPP (\ref{td-VPP}).
	\begin{equation}
		\setlength{\abovedisplayskip}{3pt}
		\setlength{\belowdisplayskip}{3pt}
		\Delta P_\text{VPP} = -2H_\text{VPP}\frac{\text{d}\Delta f}{\text{d}t}-D_\text{VPP}(\Delta f+f_\text{DB1}), t\ge t_\text{DB1}
		\label{td-VPP}
	\end{equation}
	where $H_\text{VPP}$ and $D_\text{VPP}$ are the virtual inertia and virtual damping of the VPP aggregated from numerous IBRs, $f_\text{DB1}$ is the width of the dead band of the VPP and $t_\text{DB1}$ is the time point that the VPP starts to provide droop control.   
	
	On this basis, the system frequency is supported effectively and its deviation is reformulated as in (\ref{swingnew}). 
	\begin{equation}
		\setlength{\abovedisplayskip}{3pt}
		\setlength{\belowdisplayskip}{3pt}
		\Delta P_\text{dis}+\Delta P_\text{VPP}+\Delta P_\text{PFR} = 2H_{0}\frac{\text{d}\Delta f}{\text{d}t}+D_{0}\Delta f 
		\label{swingnew}
	\end{equation}
	
	The block diagram of system frequency response is shown in Fig. \ref{fig.block} and its time-domain representation $\Delta f(t)$ is derived in (\ref{fre_2}) and the intermediate variables are listed in (\ref{variable_fre}) (the power disturbance is considered as a step drop $\Delta P_\text{dis}(s)=-\Delta P/s$ and the detail process is shown in the Appendix). $t_\text{DB1}$ and $t_\text{DB2}$ are the time points when the PFR of the VPP and SGs are activated, respectively. Due to the very short interval, the frequency response between $t_\text{DB1}$ and $t_\text{DB2}$ is omitted. 	
	
	\begin{equation}
		\Delta f(t) = \left\{ \begin{aligned}
			&- \frac{{\Delta P}}{{{D_0}}} \cdot \left( {1 - {e^{ - \frac{{{D_0}}}{{2H}} \cdot t}}} \right),0 \le t \le {t_\text{DB1}}\\
			&-\frac{{\Delta P + {D_\text{VPP}}{f_\text{DB1}}}}{{{D_\text{VPP}} + {D_0} + R}} \cdot \left[ {1 + {e^{ - \zeta {\omega _n}t}}{\eta _1}\sin ({\omega _d}t + {\varphi _1})} \right] - \\
			&\frac{{R \cdot {f_\text{DB2}}}}{{{D_\text{VPP}} + {D_0} + R}} \cdot \left[ {1 - {e^{ - \zeta {\omega _n}t}}{\eta _2}\sin ({\omega _d}t + {\varphi _2})} \right],t \ge {t_\text{DB2}}
		\end{aligned} \right.			
		\label{fre_2}		
	\end{equation}
	where
	\begin{subequations}
		\setlength{\abovedisplayskip}{3pt}
		\setlength{\belowdisplayskip}{3pt}
		\begin{align}
			& H = H_{0}+H_\text{VPP}, D = D_{0}+D_\text{VPP}\\
			& {{\omega }_{n}}=\sqrt{\frac{D+R}{2H{{T}^\text{SG}}}}, \zeta =\frac{2H+D{{T}^\text{SG}}}{2\sqrt{2{{T}^\text{SG}}H(R+D)}}\\
			& {{\omega }_{d}}={{\omega }_{n}}\sqrt{1-{{\zeta }^{2}}}, \varphi_{1} =\arctan (\frac{{{\omega }_{d}}}{-T^\text{SG}\omega _{n}^{2}+\zeta {{\omega }_{n}}})\\
			& \eta_{1} =\sqrt{\frac{1-2T^\text{SG}{{\omega }_{n}}\zeta +{{T^\text{SG}}^{2}}\omega _{n}^{2}}{1-{{\zeta }^{2}}}}\\
			& \varphi_{2}=\arctan (\frac{\sqrt{1-{{\zeta }^{2}}}}{\zeta }), \eta_{2}=\frac{1}{\sqrt{1-{{\zeta }^{2}}}} 
		\end{align}
		\label{variable_fre}
	\end{subequations}

    The analysis of this paper is based on the deterministic disturbance scenario (e.g., step changes in $\Delta P_{e}$) to analytically characterize the dynamic interplay between VPP response and reserve requirements. The deterministic disturbance assumption is based on previous works~\cite{Add-7,COM1,DER,VPP1,Add-8} and can be viewed as an approximation of ultra-short-term high-precision load/output prediction data in the real power system operation. The predicted disturbance curve can be divided at regular time intervals, and when the number of segments is sufficiently large, each segment of the disturbance can be approximated as a step signal.
    
	\section{Reserve-Minimizing Strategy for Frequency Regulation } \label{sec:strategy}
	
	\subsection{Decay Rate Constraint for VPP Interaction}
	
\begin{figure}
		\centering
		\includegraphics[width=0.4\linewidth]{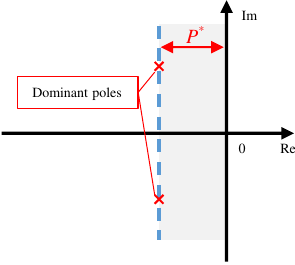}
		\caption{{Schematic diagram of the stability margin ($P^*$).}} 
		\label{fig.decay}
	\end{figure}

	The stability issue and the response speed of the VPP-integrated system depend on the dominant pole of the state matrix $\bf{A}$. Specifically, the stability margin $P^*$ refers to the magnitude of the real parts of the dominant poles as shown in Fig. \ref{fig.decay}. To make sure the system has stable and fast enough responses, the dominant poles of the state matrix $\bf{A}$ need to be on the left side of the imaginary axis with enough margin (called decay rate)~\cite{decay-1}.
	
    {However, the relationship among the stability margin $P^*$ and variables $H_\text{VPP}$ as well as $D_\text{VPP}$ is too complex to derive the analytical expressions directly. In this paper, the nonlinear surface fitting method is employed to approximate the stability margin $P^*$. The chosen form in (\ref{nonfit}) includes two linear terms ($b_{2}H_\text{VPP}$, $b_{3}D_\text{VPP}$) and a cross-term ($b_{4}H_\text{VPP}D_\text{VPP}$). }
    
	\begin{equation}
		\setlength{\abovedisplayskip}{3pt}
		\setlength{\belowdisplayskip}{3pt}
		{P^*} = {b_1} + {b_2}{H_\text{VPP}} + {b_3}{D_\text{VPP}} + {b_4}{H_\text{VPP}}{D_\text{VPP}}
		\label{nonfit}
	\end{equation}
	where $b_{1}$, $b_{2}$, $b_{3}$, and $b_{4}$ are the nonlinear fitting parameters. The linear terms capture first-order sensitivities of virtual inertia ($H_\text{VPP}$) and damping ($D_\text{VPP}$), while the cross-term accounts for the nonlinear interactions between $H_\text{VPP}$ and $D_\text{VPP}$. Specifically, the fitting parameters $b_{1}$, $b_{2}$, $b_{3}$, and $b_{4}$ are parameters to be determined by traversing different values of $P^*$ under different values of $H_\text{VPP}$ and $D_\text{VPP}$.
    
    The choice of $H_\text{VPP}$ and $D_\text{VPP}$ as variables stems from their dominant role in shaping frequency dynamics, which guide the reserve minimization and allocation discussed in the following sections. The variable $H_\text{VPP}$  governs the rate of change of frequency (RoCoF) while $D_\text{VPP}$ directly modulates the nadir values as well as quasi-steady-state (Qss) deviations. Other parameters (e.g., $T^\text{SG}$, $R$) are typically treated as accessible constants during reserve minimization and allocation. Thus, for the purpose of this research, only $H_\text{VPP}$ and $D_\text{VPP}$ are taken as variables for stability margin approximation. Thus, the decay rate constraint ($P^{*}\le \sigma < 0$) is set as in (\ref{margin}) to restrict the feasible region of $H_\text{VPP}$ and $D_\text{VPP}$.
	\begin{equation}
		\setlength{\abovedisplayskip}{3pt}
		\setlength{\belowdisplayskip}{3pt}
		{b_1} + {b_2}{H_\text{VPP}} + {b_3}{D_\text{VPP}} + {b_4}{H_\text{VPP}}{D_\text{VPP}} \le \sigma
		\label{margin}
	\end{equation}
	
	\subsection{Safety-based Constraints for System Frequency}	
	
	\begin{figure}
		\centering
		\includegraphics[width=0.8\linewidth]{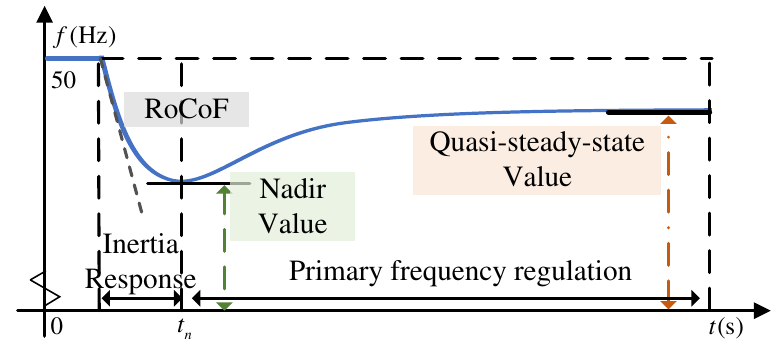}
		\caption{The system frequency response can be divided into inertia response and primary frequency regulation.} 
		\label{fig.response}
	\end{figure}
	
	As shown in Fig. \ref{fig.response}, based on the time-domain expression of system frequency response in (\ref{fre_2}), the system frequency metrics can be derived, which include RoCoF in (\ref{con.1}), the nadir value in (\ref{con.2})-(\ref{con.2-1}) and the Qss of frequency deviation in (\ref{con.3}).
	\begin{equation}
		\setlength{\abovedisplayskip}{3pt}
		\setlength{\belowdisplayskip}{3pt}
		\Delta f_{\max }^\text{RoCoF}={{\left. \Delta \dot{f}(t) \right|}_{t={{0}^{+}}}}=-\frac{\Delta P}{2(H_{0}+H_\text{VPP})}
		\label{con.1}
	\end{equation}
	
	\begin{equation}
		\setlength{\abovedisplayskip}{3pt}
		\setlength{\belowdisplayskip}{3pt}
		{\left. \Delta {{f}^\text{Nadir}} \right|}_{\Delta \dot{f}(t_{n})=0} = \Delta f({{t}_{n}})\\
		\label{con.2}
	\end{equation}
	where $t_{n}$ is the time that frequency deviation reaches its lowest point (nadir value), ${{t}_{n}}=\frac{\arctan (N)}{{{\omega }_{d}}}$.

	\begin{equation}
		\setlength{\abovedisplayskip}{3pt}
		\setlength{\belowdisplayskip}{3pt}
		N =\frac{{{\omega }_{d}}(m\cos ({\varphi_{2}})-\cos (\varphi_{1}))-\zeta {{\omega }_{n}}(m\sin ({\varphi_{2}})+\sin (\varphi_{1}))}{\zeta {{\omega }_{n}}(m\cos({\varphi_{2}})-\cos (\varphi_{1}))+{{\omega }_{d}}(m\sin ({\varphi_{2}})-\sin (\varphi_{1}))}
	\end{equation}
	\begin{equation}
		\setlength{\abovedisplayskip}{3pt}
		\setlength{\belowdisplayskip}{3pt}
		m=\frac{R{{f}_\text{DB2}}}{\Delta P+{{D}_\text{VPP}}{{f}_\text{DB1}}}\cdot \frac{{{\eta_{2}}}}{\eta_{1}}
        \label{con.2-1}
	\end{equation}
	\begin{equation}
		\setlength{\abovedisplayskip}{3pt}
		\setlength{\belowdisplayskip}{3pt}
		\Delta {{f}^\text{Qss}}={{\left. \Delta f(t) \right|}_{t\to \infty }}=-\frac{\Delta P+{{D}_\text{VPP}}{{f}_\text{DB1}}+R\cdot {{f}_\text{DB2}}}{{{D}_\text{VPP}}+{{D}_{0}}+R}
		\label{con.3}
	\end{equation}
	
	\subsection{Feasible Region of Regulation Requirements for VPP}
	
	As for the interaction of VPP, the capacity constraints for transmission lines should be taken into consideration for determining the upper bounds of $H_\text{VPP}$ and $D_\text{VPP}$. The power in various transmission lines is time-varying and can be calculated by the power transfer distribution factor~(PTDF) of the grid as in (\ref{linecapacity}).
	\begin{equation}
		\setlength{\abovedisplayskip}{3pt}
		\setlength{\belowdisplayskip}{3pt}
		\left|{s^{l}\Delta P_{\text{VPP},t}}+\sum\nolimits_{g}{s_{g}^{l}P_{g,t}^\text{SG}}-\sum\nolimits_{d}{s_{d}^{l}{{L}_{t}}}\right| \le P^{lm},\forall l,t	
		\label{linecapacity}
	\end{equation}
	where $P^{lm}$ represent the upper limits of transmission lines' capacities, $s_{i}^{l}, s_{g}^{l}, s_{d}^{l}$ represent the PTDF of transmission line $l$ \cite{COM2}. 
	The analytical formulations of $\Delta P_\text{VPP}$ is derived in (\ref{vpp_power})-(\ref{vpp_power_d}). Thus, the limit of power injections from the VPP ($\Delta P_\text{VPP}$) can be transferred into bounds of $H_\text{VPP}$ and $D_\text{VPP}$.	
	
	\begin{equation}
		\setlength{\abovedisplayskip}{3pt}
		\setlength{\belowdisplayskip}{3pt}
		\Delta P_\text{VPP}(t) = \Delta P_\text{VPP}^s(t)+\Delta P_\text{VPP}^d(t)
		\label{vpp_power}		
	\end{equation}	
	where $\Delta P_\text{VPP}^s(t)$ and $\Delta P_\text{VPP}^d(t)$ are the static component and dynamic component of power injections from the VPP, respectively.
	\begin{equation}
		\setlength{\abovedisplayskip}{3pt}
		\setlength{\belowdisplayskip}{3pt}
		\Delta P_\text{VPP}^s(t) = \frac{A-2H{{T}^\text{SG}}\cdot \Delta {{P}_{3}}}{2H{{T}^\text{SG}}}
		\label{vpp_power_s}
	\end{equation}
	
	\begin{equation}
		\Delta P_\text{VPP}^d(t) = \frac{{{e}^{-\zeta {{\omega }_{n}}t}}\left( \frac{C-\zeta {{\omega }_{n}}\cdot B}{{{\omega }_{d}}}\sin ({{\omega }_{d}}t)+B\cos ({{\omega }_{d}}t) \right) }{2H{{T}^\text{SG}}}
		\label{vpp_power_d}		
	\end{equation}
	where $A=\frac{(\Delta {{P}_{1}}+\Delta {{P}_{2}}){{D}_\text{VPP}}}{\omega _{n}^{2}}$, $B=2{{T}^\text{SG}}{{H}_\text{VPP}}\Delta {{P}_{1}}-\frac{(\Delta {{P}_{1}}+\Delta {{P}_{2}}){{D}_\text{VPP}}}{\omega _{n}^{2}}$, $C={{T}^\text{SG}}{{D}_\text{VPP}}\cdot$
	$\Delta {{P}_{1}}+2{{H}_\text{VPP}}(\Delta {{P}_{1}}+\Delta {{P}_{2}})-\frac{2\zeta (\Delta {{P}_{1}}+\Delta {{P}_{2}}){{D}_\text{VPP}}}{{{\omega }_{n}}}$, $\Delta {{P}_{1}}=\Delta P+{{D}_\text{VPP}}{{f}_\text{DB1}}$, $\Delta {{P}_{2}}=R\cdot {{f}_\text{DB2}}$, $\Delta {{P}_{3}}={{D}_\text{VPP}}{{f}_\text{DB1}}$.
	
	To guarantee the safety of the grid's transmission lines, the virtual inertia and damping from the VPP cannot go beyond a certain limit $H_\text{VPP}^\text{max}$ and $D_\text{VPP}^\text{max}$. Therefore, the closed-form representation of the feasible region for the regulation requirement ($H_\text{VPP}$, $D_\text{VPP}$) is derived in (\ref{safety}).
	\begin{equation}
		\left\{ \begin{aligned}
			& H_\text{VPP} \ge \frac{\Delta P}{2\Delta f_{\lim }^\text{RoCoF}}-H_{0}  \\ 
			& \Delta f(t_{n}) \le -\Delta f_{\lim }^\text{Nadir} \\ 
			& D_\text{VPP} \ge \frac{\Delta P+R f_\text{DB2}-\Delta f_{\lim }^\text{Qss}(D_{0}+R)}{\Delta f_{\lim }^\text{Qss}-f_\text{DB1}}  \\ 
			& {b_1} + {b_2}{H_\text{VPP}} + {b_3}{D_\text{VPP}} + {b_4}{H_\text{VPP}}{D_\text{VPP}} \le \sigma\\
			& 0 \le H_\text{VPP} \le H_\text{VPP}^{\max}, 0 \le D_\text{VPP} \le D_\text{VPP}^{\max}\\
		\end{aligned} \right.
		\label{safety}
	\end{equation}
	
	\subsection{Minimal Reserve Decision of VPP}
	In this section, we propose a strategy to determine the minimal reserve requirements for the VPP. Specifically, the second-level reserves refer to the active power injection requirements necessary to maintain frequency stability. According to equations (\ref{vpp_power}) - (\ref{vpp_power_d}), the active power injections can be determined by the parameter combinations ($H_\text{VPP}$, $D_\text{VPP}$) within the multi-constraint feasible region. 	
	
	The steady-state value of power injection can be calculated as in (\ref{steady_value}), where $\Delta P_{{\text{VPP}}}(t \to \infty )$ is in direct proportion to $D_\text{VPP}$. The steady-state value of power injection $\Delta P_{{\text{VPP}}}(t \to \infty )$ dominates the cumulative energy injection \cite{COM1}. Thus, we first tune $D_\text{VPP}$ to its minimum ($D_\text{VPP}^{re}$) to cut extra amounts of power injections. 
	\begin{equation}
		\setlength{\abovedisplayskip}{3pt}
		\setlength{\belowdisplayskip}{3pt}
		\begin{aligned}
			\Delta P_{{\text{VPP}}}(t \to \infty ) = \frac{\Delta P + R{f_\text{DB2}} - (D_{0}+R){f_\text{DB1}}}{1+\frac{D_{0}+R}{D_\text{VPP}}}
		\end{aligned}
		\label{steady_value}
	\end{equation}
	
	For determining aggregated virtual inertia ($H_\text{VPP}$), we derive the cumulative energy injection of VPP during the period of frequency support in (\ref{int_power}). 
	\begin{equation}
		\setlength{\abovedisplayskip}{3pt}
		\setlength{\belowdisplayskip}{3pt}
		\begin{aligned}
			E = \int_0^T &{\Delta {P_\text{VPP}}(t){\rm{d}}t}  = \frac{{A - 2H{T^\text{SG}} \cdot \Delta {P_3}}}{{2H{T^\text{SG}}}}T \\
			&+ \frac{{\int_0^T {{e^{ - \zeta {\omega _n}t}}\left( {\frac{{C - \zeta {\omega _n}B}}{{{\omega _d}}}\sin ({\omega _d}t) + B\cos ({\omega _d}t)} \right){\rm{d}}t} }}{{2H{T^\text{SG}}}}
		\end{aligned}
		\label{int_power}
	\end{equation}
	where $T$ is the regulation time. Based on the assumption that $e^{ - \zeta {\omega _n}T}\approx0$, we simplify the integration as in (\ref{int_app}). Substitute the (\ref{int_app}) into (\ref{int_power}) to derive the differential expression for $H_\text{VPP}$ in (\ref{int_de}).
	\begin{equation}
		\begin{aligned}
			&\int_0^T {{e^{ - \zeta {\omega _n}t}}\left( {\frac{{C - \zeta {\omega _n}B}}{{{\omega _d}}}\sin ({\omega _d}t) + B\cos ({\omega _d}t)} \right){\rm{d}}t} \\
			&\approx \frac{{C - \zeta {\omega _n}B}}{{{\omega _d}}} \cdot \frac{{{\omega _d}}}{{{{(\zeta {\omega _n})}^2} + \omega _d^2}} + \frac{{B\zeta {\omega _n}}}{{{{(\zeta {\omega _n})}^2} + \omega _d^2}} = \frac{C}{{\omega _n^2}}
		\end{aligned}
		\label{int_app}
	\end{equation}
	
	\begin{equation}
		\begin{aligned}
			\frac{{{\rm{d}}\left( {\int_0^T {\Delta {P_\text{VPP}}(t)\rm{d}t} } \right)}}{{{\rm{d}}{H_\text{VPP}}}} \approx \frac{{2(\Delta {P_1} + \Delta {P_2})}}{{{{(D + R)}^2}}}({D_0} + R) > 0
		\end{aligned}
		\label{int_de}
	\end{equation}
	
	According to (\ref{int_de}), the energy of VPP supporting the system frequency is in direct proportion with $H_\text{VPP}$ under the determination of $D_\text{VPP}^{re}$. Thus, the specific requirement for $H_\text{VPP}^{re}$ is tuned to minimize the energy injection in (\ref{minH}). 
	\begin{equation}
		\begin{aligned}
			& H_\text{VPP}^{re} = \text{argmin}\ E(H_\text{VPP} ,D_\text{VPP})\\
			& s.t.\\
			& D_\text{VPP} = D_\text{VPP}^{re}, (\ref{safety})
		\end{aligned}
		\label{minH}
	\end{equation}
	
	In this way, the required virtual inertia as well as damping can be determined sequentially and is completely described in Algorithm \ref{alg.1}. With the optimal combination of inertia and damping, the minimal reserve requirements are determined as well for the VPP.  
	\begin{figure}[h]
		\renewcommand{\algorithmicrequire}{\textbf{Input:}}
		\renewcommand{\algorithmicensure}{\textbf{Output:}}
		\begin{algorithm}[H]
			\caption{Two-Stage Algorithm for Minimal Reserve Decision of VPP}
			\begin{algorithmic}[1]
				\REQUIRE The grid conditions~($H_{0}$, $D_{0}$, $\Delta P$) and corresponding metrics~($\Delta f_\text{lim}^\text{RoCoF}$, $\Delta f_\text{lim}^\text{Nadir}$, $\Delta f_\text{lim}^\text{Qss}$, $\sigma$).        	%%input
				\ENSURE  The required aggregated virtual inertia $H_\text{VPP}^{re}$ and damping $D_\text{VPP}^{re}$ for VPP.  		%%output	
				\STATE{Derive the multi-constraint feasible region of $H_\text{VPP}$ and $D_\text{VPP}$ in (\ref{safety}).}		
				\STATE {Calculate the required $D_\text{VPP}^{re}$ that ensures the minimal $\Delta P^\text{VPP}(t \to \infty)$ in (\ref{steady_value}) in the feasible region.}
				\STATE {Based on the determined $D_\text{VPP}^{re}$, calculate the required $H_\text{VPP}^{re}$ in (\ref{minH}) that ensures the minimal $E$.}
			\end{algorithmic}
			\label{alg.1}
		\end{algorithm}
	\end{figure}
	
	\begin{figure}
		\centering
		\includegraphics[width=0.7\linewidth]{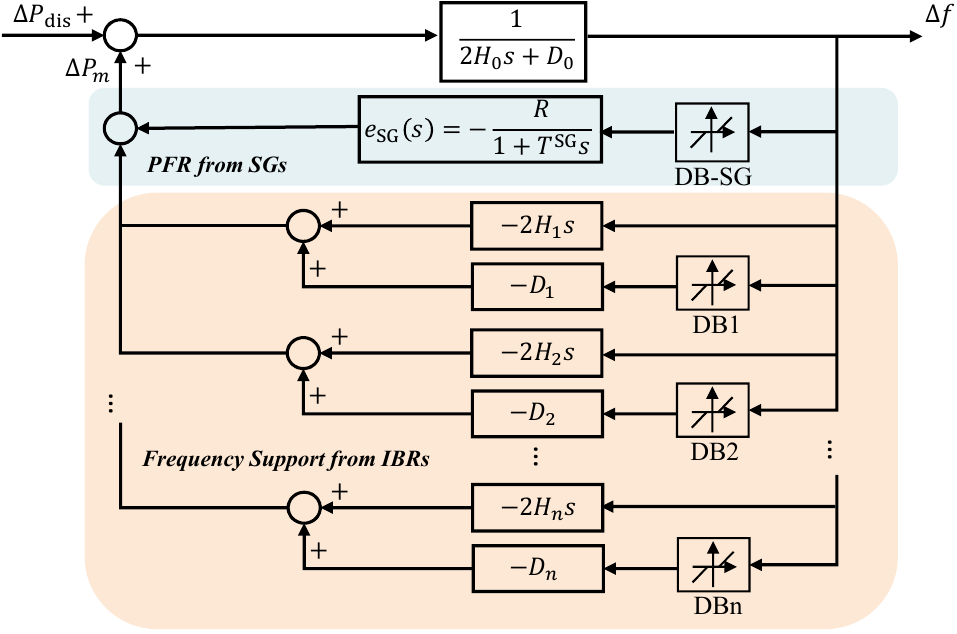}
		\caption{The equivalent control diagrams of VPP-level and IBR-level frequency support based on the allocation of virtual inertia and damping requirements.} 
		\label{fig.disagg}
	\end{figure}
	
	\section{Optimal Allocation for IBR-level Reserves}	\label{sec:allocation}
	
	Based on the determined $H_\text{VPP}^{re}$ and $D_\text{VPP}^{re}$, the corresponding regulation reserve for VPP is obtained, which should be allocated to IBRs totally. According to the disaggregation strategy in Fig. \ref{fig.disagg}, the active power injections of the $i$th IBR are derived in (\ref{ibr_power}) under the assumption that the widths of IBRs' dead bands are the same.
	\begin{equation}
		\Delta P_{\text{IBR},i}(t) = \Delta P_{\text{IBR},i}^s(t)+\Delta P_{\text{IBR},i}^d(t)
		\label{ibr_power}		
	\end{equation}	
	where $\Delta P_{\text{IBR},i}^s(t)$ and $\Delta P_{\text{IBR},i}^d(t)$ are the static and dynamic components of power injections from the $i$th IBR, respectively.
	\begin{equation}
		\Delta P_{\text{IBR},i}^s(t) = \frac{\mathbb{A}-2H{{T}^\text{SG}}\cdot \Delta {\mathbb{P}_{3}}}{2H{{T}^\text{SG}}}
		\label{ibr_power_s}
	\end{equation}
	
	\begin{equation}
		\Delta P_{\text{IBR},i}^d(t) = \frac{{{e}^{-\zeta {{\omega }_{n}}t}}\left( \frac{\mathbb{C}-\zeta {{\omega }_{n}}\cdot \mathbb{B}}{{{\omega }_{d}}}\sin ({{\omega }_{d}}t)+\mathbb{B}\cos ({{\omega }_{d}}t) \right) }{2H{{T}^\text{SG}}}
		\label{ibr_power_d}		
	\end{equation}
	where $\mathbb{A}=\frac{(\Delta {{P}_{1}}+\Delta {{P}_{2}}){{D}_{i}}}{\omega _{n}^{2}}$, $\mathbb{B}=2{{T}^\text{SG}}{{H}_{i}}\Delta {{P}_{1}}-\frac{(\Delta {{P}_{1}}+\Delta {{P}_{2}}){{D}_{i}}}{\omega _{n}^{2}}$, $\mathbb{C}={{T}^\text{SG}}{{D}_{i}}\Delta {{P}_{1}}+2{{H}_{i}}(\Delta {{P}_{1}}+\Delta {{P}_{2}})-\frac{2\zeta (\Delta {{P}_{1}}+\Delta {{P}_{2}}){{D}_{i}}}{{{\omega }_{n}}}$, $\Delta {{P}_{1}}=\Delta P+{{D}_\text{VPP}}{{f}_\text{DB1}}$, $\Delta {{P}_{2}}=R\cdot {{f}_\text{DB2}}$, $\Delta {\mathbb{P}_{3}}={{D}_{i}}{{f}_\text{DB1}}$. 
	
	The aim is to maximize the financial profits of the VPP that provides frequency support by setting specific regulation reserves. These reserves are economically compensated by the power grid~(\ref{financial_R}). The economic diversity of IBRs results in varying costs due to different reserve allocation strategies~(\ref{financial_C}). 
	\begin{equation}
		\setlength{\abovedisplayskip}{3pt}
		\setlength{\belowdisplayskip}{3pt}
		F_{\text{Res}} = {c^F}\sum\nolimits_{i=1}^{N^\text{IBR}}\sum\nolimits_{t = 1}^T \Delta P_{\text{IBR},i}(t)
		\label{financial_R}
	\end{equation}
	\begin{equation}
		F_{\text{Cost}} = -\sum\nolimits_{i=1}^{N^\text{IBR}} c_{i}^{R}\sum\nolimits_{t=1}^{T} \Delta P_{\text{IBR},i}(t)
		\label{financial_C}
	\end{equation}
	where $N^\text{IBR}$ is the number of IBRs involved in the VPP, $c^{F}$ is the unit profit for providing reserve, $c_{i}^{R}$ is the unit cost of active power injections from the $i$th IBR, $c^{P}$ is the unit unsatisfied punishment for frequency support, $P_{i}^\text{rated}$ represents the upper limits of IBR's regulation power capacity, $T$ is the time duration of the frequency regulation, $\Delta P^{re}_{\text{VPP}}$ and $\Delta P^{ac}_{\text{VPP}}$ are the reserve requirements from the power grid and the actual reserve provided by the VPP respectively. Thus, the optimization model is established in (\ref{model_opt}).
	\begin{align}
		\label{model_opt}
		& {\underset{{{H}_{i}},{{D}_{i}}}{\mathop{\max }} \ F = F_{\text{Res}}+F_{\text{Cost}} }\notag\\
		& {s.t.}  \notag \\ 
		& {\sum\nolimits_{i=1}^{N^\text{IBR}}{{{H}_{i}}}={{H}_\text{VPP}^{re}},} \notag\\ 
		& {\sum\nolimits_{i=1}^{N^\text{IBR}}{{{D}_{i}}}={{D}_\text{VPP}^{re}},} \\ 
		& {{{H}^{\min }_\text{IBR}}\le {{H}_{i}}\le {{H}^{\max }_\text{IBR}}, \forall i\in N^\text{IBR}} \notag\\
		& {{{D}^{\min }_\text{IBR}}\le {{D}_{i}}\le {{D}^{\max }_\text{IBR}}, \forall i\in N^\text{IBR}}  \notag \\ 
		& {0\le \Delta P_{\text{IBR},i}(t)\le  P_{i}^\text{rated}, \forall t\in T, \forall i\in N^\text{IBR}}  \notag
	\end{align}
	where $H_\text{VPP}^{re}$ and  $D_\text{VPP}^{re}$ are determined based on Algorithm~\ref{alg.1} for minimal reserve decision. It can be seen that the objective function is dominated by the active power injection of the $i$th IBR and it can be decomposed and expanded as in (\ref{power}).
	\begin{equation}
		\setlength{\abovedisplayskip}{3pt}
		\setlength{\belowdisplayskip}{3pt}
		\begin{aligned}
			& \Delta P_{\text{IBR},i}(t) = \Delta P_{\text{IBR},i}^s(t)+\Delta P_{\text{IBR},i}^d(t)	\\
			& ={({k}^{D}+\beta)}\cdot {{D}_{i}}+\alpha \cdot {{H}_{i}}
		\end{aligned}	
		\label{power}
	\end{equation}
	where $k^{D}$, $\alpha$ and $\beta$ are the intermediate variables and their detailed representations are listed in (\ref{linearpara}). The variables depend on system-level parameters and can be considered as constants for $H_{i}$ and $D_{i}$. Thus, the contributions of $H_{i}$ and $D_{i}$ to the active power injections are decoupled, which can guide the following economic analysis. What's more, as for the variables $H_{i}$ and $D_{i}$, the model (\ref{model_opt}) is a linear as well as convex program, which can be directly solved to its optimum.
	\begin{equation}
		\begin{aligned}
			&{k^D} = \left( {\frac{{\Delta {P_1} + \Delta {P_2}}}{{\omega _n^2}} - 2H{T^\text{SG}}{f_\text{DB1}}} \right) \cdot \frac{1}{{2H{T^\text{SG}}}}\\
			&\alpha  = \frac{{{e^{ - \zeta {\omega _n}t}}\sin ({\omega _d}t)}}{{H{\omega _d}}}\left( {\frac{{\Delta {P_1} + \Delta {P_2}}}{{{T^{{\rm{SG}}}}}} - \zeta {\omega _n}\Delta {P_1}} \right) + \frac{{{e^{ - \zeta {\omega _n}t}}\cos ({\omega _d}t)}}{H} \Delta P_{1}\\
			&\beta  = \frac{{{e^{ - \zeta {\omega _n}t}}\cos ({\omega _d}t)(\Delta {P_1} + \Delta {P_2})}}{{2H{T^{{\rm{SG}}}}\omega _n^2}} + \frac{{{e^{ - \zeta {\omega _n}t}}\sin ({\omega _d}t)}}{{2H{\omega _d}}}\left[ {\Delta {P_1} - \frac{{\zeta (\Delta {P_1} + \Delta {P_2})}}{{{T^{{\rm{SG}}}\omega_n}}}} \right]
		\end{aligned}
		\label{linearpara}
	\end{equation}

    The proposed reserve strategy, as in~(\ref{model_opt}), faces inherent risks of unsatisfied frequency support due to the uncertainty of renewable energy and possible model errors. Actually, the risks always exist because the capacity of battery storage is often limited in real power system operations. The primary source of uncertainty arises from renewable energy, such as wind and solar power, which mainly affects the upper limit of the $i$th IBR’s regulation power capacity ($P_{i}^\text{rated}$), thereby restricting the active power injections ($\Delta P_{\text{IBR},i}$) based on the actual generation conditions of the $i$th IBR. Additionally, model errors may arise due to variations in parameters from the control loops of IBRs, which may also influence the active power injections ($\Delta P_{\text{IBR},i}$) quantification as well as frequency support performance of the VPP.

    To make the research more complete, we formulate the economic punishment as in (\ref{financial_P}) to evaluate the impacts of possible uncertainty, where $\Delta P_\text{VPP}^{re}(t)$ and $\Delta P_\text{VPP}^{ac}(t)$ are the reserve requirements from the power grid and the actual reserve provided by the VPP respectively.
    
	\begin{equation}
		{F_{{\rm{Pun}}} =  - {c^P}\sum\nolimits_{t = 1}^T {\left( {\Delta P_{{\rm{VPP}}}^{re}(t) - \Delta P_{{\rm{VPP}}}^{ac}(t)} \right)}} 
		\label{financial_P}
	\end{equation}

	\begin{figure}
		\centering
		\includegraphics[width=0.8\linewidth]{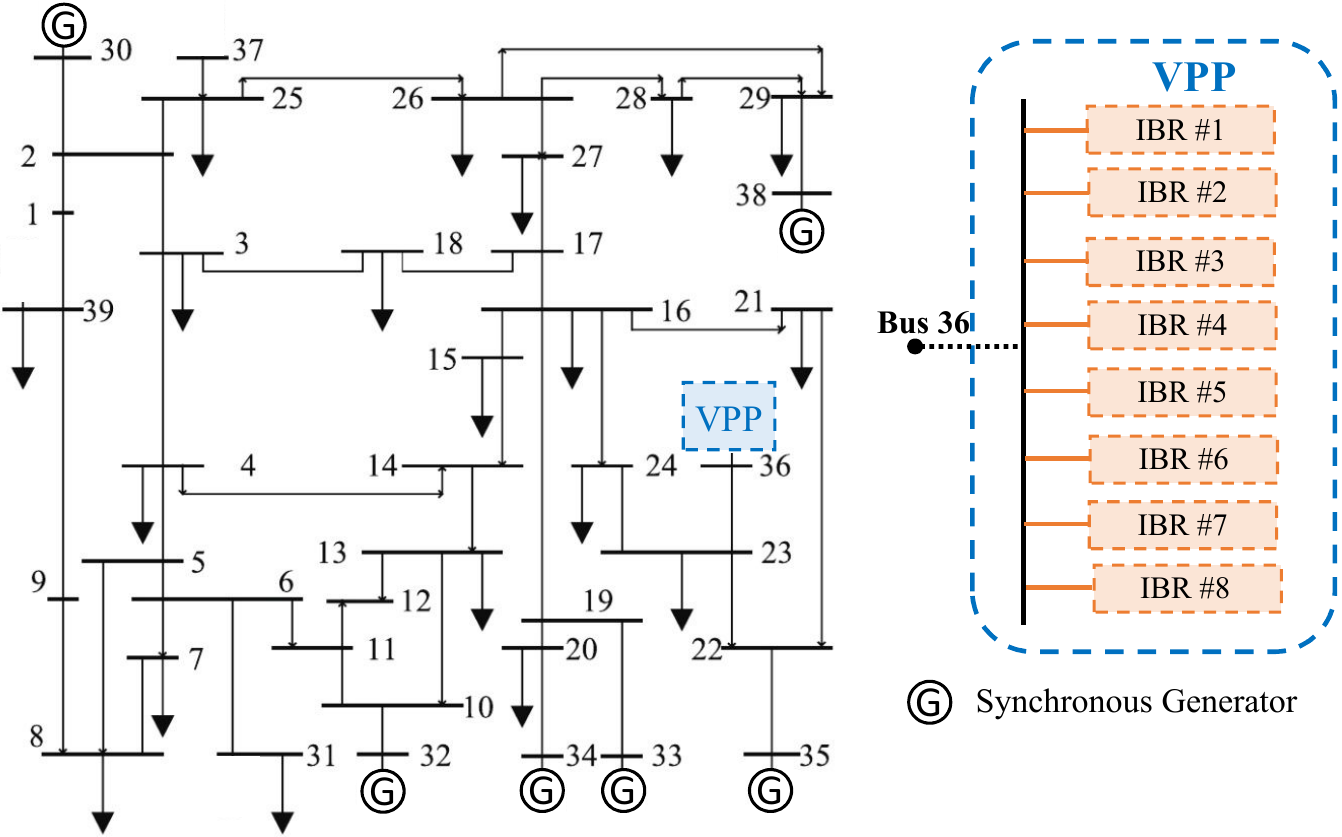}
		\caption{The modified IEEE-39 Bus topology for case studies.} 
		\label{fig.topo}
	\end{figure}
	
	\section{Case Studies} \label{sec:cases}
	
	\subsection{Simulation Setup}
	
	The proposed strategy is verified on a modified IEEE-39bus system with eight IBR-interfaced resources (in Fig. \ref{fig.topo}). The base power, frequency and voltage are set as 1000 MVA, 50 Hz and 690 V and other system parameter values used for cases are listed in Table \ref{tab:parameter} \cite{COM1}. Case studies are conducted on MATLAB and Gurobi with a desktop with IntelCore i7-10700 2.90GHz CPU.
	
	As for the regulator data for VPP, the parameters are set as $i_{0}^{V}=0.25$ (p.u.), $K_{p}^{P}=0.637$, $K_{p}^{I}=6.37$, $K_{r}^{P}=20000$, $K_{r}^{I}=150000$, $L_{1}=11.4$ (mH), $R_{1} = 3.57$ ($\Omega$), $f_{0} = 50$ (Hz), $f_\text{DB1} = 0.03$ (Hz), $f_\text{DB2} = 0.033$ (Hz). As for the safety constraints of system frequency, the limit of RoCoF $\Delta f_\text{lim}^\text{RoCoF}$ is set $0.4$ (Hz/s), the limit of nadir $\Delta f_\text{lim}^\text{Nadir}$ is set $0.5$ (Hz), the limit of Qss $\Delta f_\text{lim}^\text{Qss}$ is set $0.35$ (Hz). As for the decay rate constraint, the parameter $\sigma$ is set $-0.3$.
	
	\begin{table}[htbp]
		\centering
		\caption{System parameters for case studies}
		\begin{tabular}{ccc}
			\toprule
			Parameter & Value & Unit \\
			\midrule
			$D_{0}$ & 2     & p.u. \\
			$H_{0}$ & 5     & s \\
			$R$ & 25    & \textbackslash{} \\
			$T^\text{SG}$ & 5     & s \\
			$N^\text{IBR}$ & 8     & \textbackslash{} \\
			$\Delta P_{e}$ & 0.25   & p.u. \\
			\multirow{2}[0]{*}{$c_{i}^{R}$} & 20.61,18.96,19.15,20.06, & \multirow{2}[0]{*}{\$/MWh} \\
			& 19.15,20.61,18.96,20.06 &  \\
			{$c^{F}$,$c^{P}$} & {30,90}    &  {\$/MWh} \\
			\multirow{2}[0]{*}{$P_{i}^\text{rated}$} & 0.03,0.055,0.04,0.02, & \multirow{2}[0]{*}{p.u.} \\
			& 0.01,0.06,0.02,0.015 &  \\
			$T$ & 60    & s \\
			$P^{lm}$ & 0.2    & p.u. \\
			$P^{G}_{ref}$ & 0.75    & p.u. \\
			$H^\text{min}_\text{IBR}, H^\text{max}_\text{IBR}$ & 0.1,6  & s \\
			$D^\text{min}_\text{IBR}, D^\text{max}_\text{IBR}$ & 0.1,6  & p.u. \\
			\bottomrule
		\end{tabular}%
		\label{tab:parameter}%
	\end{table}%
	
	\subsection{Comparisons of Reserve Decision for VPP}
	
\begin{figure}
		\centering
		\includegraphics[width=0.8\linewidth]{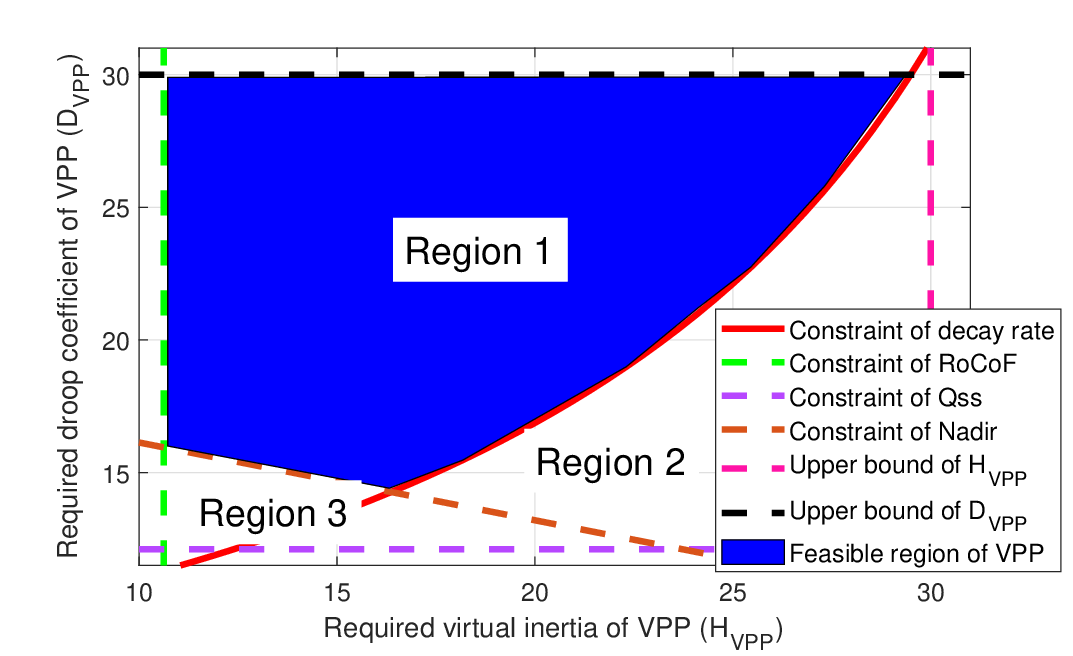}
		\caption{The multi-constraint feasible region of $H_\text{VPP}$ and $D_\text{VPP}$, that is the blue part~(Region 1).}
		\label{fig.timedom1}
	\end{figure}

	Considering the capacity limit of transmission lines of the modified IEEE-39bus system, the upper bounds of aggregated virtual inertia and damping are set as $H_\text{VPP}^\text{max}=30$ (s) and $D_\text{VPP}^\text{max}=30$ (p.u.). Under the bounds, the maximum power flow of the grid topology during frequency regulation is 0.19 (p.u.), satisfying the transmission line constraint. {As for the nonlinear surface fitting presented in Section~\ref{sec:strategy}.1, the fitting parameters are [$b_1$=-0.146, $b_2$=0.0012, $b_3$=-0.0195, $b_4$=0.0004] after convergency, with an average fitting accuracy of 96.8\%, which is acceptable for engineering applications.} On this basis, the feasible region of regulation requirements of the VPP is represented as the blue part in Fig. \ref{fig.timedom1}. 
	
	According to the two-stage algorithm for the minimal reserve decision of VPP in Algorithm \ref{alg.1}~(ResMin for short), the aggregated virtual inertia and damping for the VPP is determined as $H_\text{VPP}$=15.925 (s), $D_\text{VPP}$=14.2094 (p.u.). Thus, the minimal reserve for the VPP is determined correspondingly. Under this circumstance, the maximum RoCoF, nadir as well as Qss values of system frequency response are 0.3~(Hz/s), 0.5~(Hz) and 0.33~(Hz), respectively. It validates the safety of frequency response according to the set boundaries.  
	
	\begin{figure}
		\centering
		\includegraphics[width=0.8\linewidth]{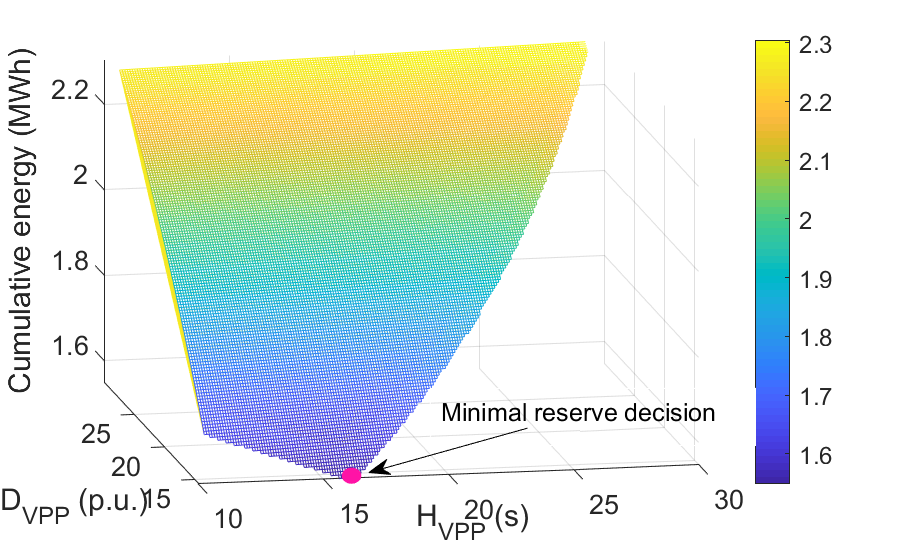}
		\caption{Cumulative energy of different combinations of virtual inertia~($H_\text{VPP}$) and virtual damping~($D_\text{VPP}$).}
		\label{fig.energy}
	\end{figure}
	
	As shown in Fig. \ref{fig.energy}, the decision obtained by ResMin~(the purple point) calls for minimal cumulative energy for frequency regulation reserve compared with other combinations~($H_\text{VPP}$, $D_\text{VPP}$). As for the comparison with the reserve decision strategy based on peak value~(ResPeak for short) \cite{COM1}, the cumulative energy requirement of ResMin is 1.54 (MWh) for reserve while that of ResPeak is 3.2 (MWh) under the same combination of $H_\text{VPP}$=15.925 (s), $D_\text{VPP}$=14.2094 (p.u.). ResMin shows a great improvement (by 51.88\%) in financial gains by precise calculation and releasing idle reserves.  
	
\begin{figure}
		\centering
		\includegraphics[width=0.8\linewidth]{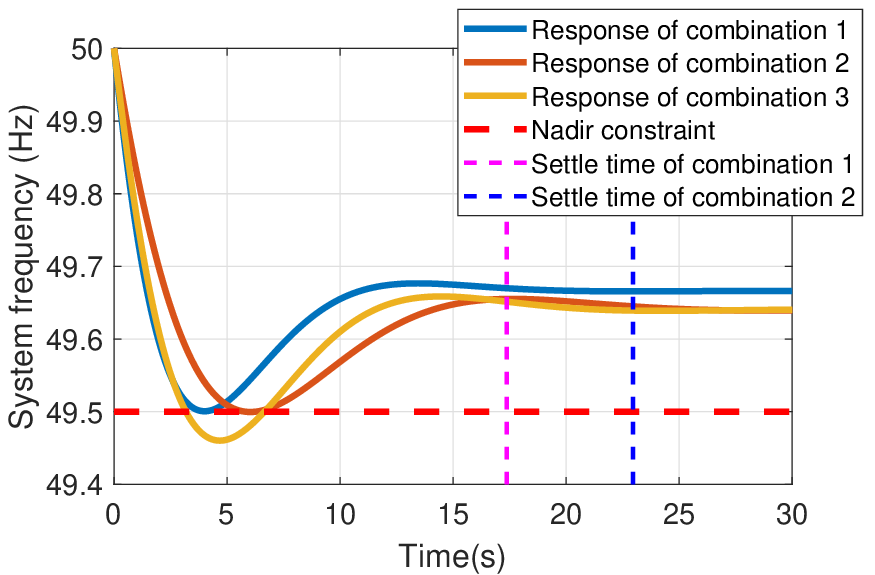}
		\caption{Time domain frequency responses of various combinations selected from three regions in Fig. \ref{fig.timedom1}.}
		\label{fig.timedom2}
	\end{figure}

	To illustrate the effectiveness of the feasible region (the blue part in Fig. \ref{fig.timedom1} marked as Region 1), we compared the time-domain frequency responses of three regions in Fig. \ref{fig.timedom2}. The selected combinations are listed in Table \ref{tab:metric}, where the combination selected from Region 1 has a safe enough nadir value with a shorter settle time. The settling time of frequency response derived from Region 1 decreases those of Region 2 and Region 3 by 24.35\% and 10.28\% due to the consideration of the decay rate.
	
	\begin{table}[htbp]
		\centering
		\caption{Comparison of time-domain metrics}
		\begin{tabular}{ccccc}
			\toprule
			Region & $H_\text{VPP}$ (s) & $D_\text{VPP}$ (p.u.) & Nadir (Hz) & Settle time (s) \\
			\midrule
			Region 1 & 15.925 & 14.2094 & \textbf{49.50} & \textbf{17.37} \\
			Region 2 & 28    & 11    & 49.50 & 22.96 \\
			Region 3 & 19    & 11    & 49.46 & 19.36 \\
			\bottomrule
		\end{tabular}
		\label{tab:metric}
	\end{table}

	\subsection{Case I: Deterministic Reserve Allocation without Considering Uncertainty}
	
\begin{figure}
		\centering
		\includegraphics[width=0.7\linewidth]{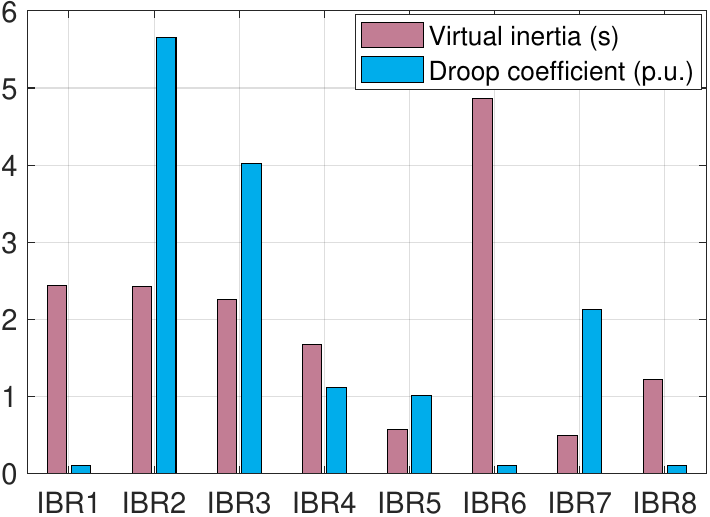}
		\caption{Optimal allocation results of virtual inertia and damping parameters for IBRs based on~(\ref{model_opt}).}
		\label{fig.case1}
	\end{figure}

	In this section, the uncertainty is not considered, which means that the power dynamics of IBRs can be accurately predicted; thus, the allocation optimization can be seen as a deterministic program. The convex optimal allocation model (\ref{model_opt}) is solved by Gurobi and the results are presented in Fig. \ref{fig.case1}. It can be seen that the values of $H_{i}$ and $D_{i}$ are varied among eight IBRs, reflecting the impacts of IBR's device-specific limitation as well as economic difference on allocation results. 
    
	The maximum power injection capacity of the $i$th IBR, constrained by $P_{i}^\text{rated}$, is significantly influenced by its virtual inertia parameter $H_{i}$. While the cumulative energy capacity of the $i$th IBR, constrained by $c_{i}^{R}$, is notably affected by its virtual damping $D_{i}$. Given that larger capacities enable better system frequency support, the 6th IBR is inclined to receive higher allocations of virtual inertia. In the meantime, the 2nd IBR, the 3rd IBR and the 7th IBR demonstrate tendencies for higher virtual damping during power injection, indicative of their heightened economic significance.

	As for the economic efficiency of reserve allocation, the comparison is conducted among the strategies of the proposed optimal allocation~(AllocOpt for short), the even allocation~(AllocEven for short) and the proportional allocation~(AllocProp for short). The description of the above three allocation methods is listed as follows.
	
	\begin{enumerate}
		\item AllocOpt: The actual reserves of IBRs are determined by optimal allocation of $H_\text{VPP}$ and $D_\text{VPP}$ according to their capacity constraints and economic preferences.
		
		\item AllocEven: The actual reserves of IBRs are determined by even allocation of $H_\text{VPP}$ and $D_\text{VPP}$ among various IBRs.
		
		\item AllocProp: The actual reserves of IBRs are determined by the proportional allocation of $H_\text{VPP}$ and $D_\text{VPP}$, based on their power capacities~\cite{DER}.
	\end{enumerate}
	
	The corresponding economic benefit of the proposed strategy~(AllocOpt) is \$17.09, compared with \$16.29 for AllocEven and \$16.21 for AllocProp during the regulation period. AllocOpt increases AllocEven and AllocProp by 4.91\% and 5.43\% due to the optimal allocation modeling and solution. The comparison results are related to the parameters set for the case studies, but the financial superiority of the proposed strategy remains unchanged. 	
		
	\begin{figure}[h]
		\centering
		\includegraphics[width=0.7\linewidth]{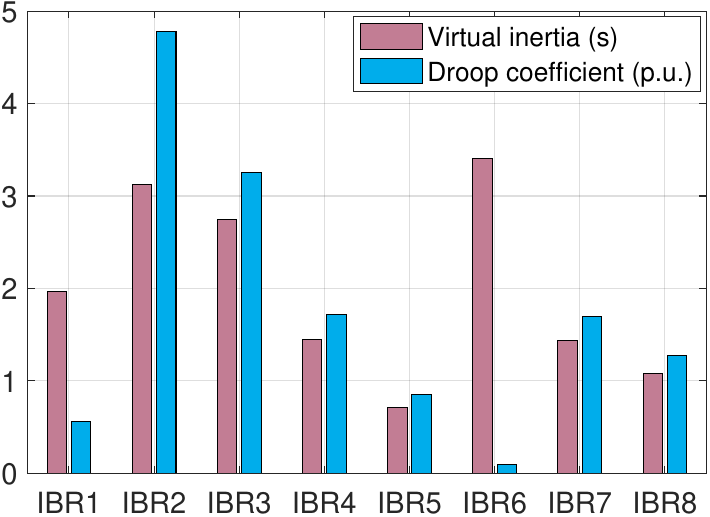}
		\caption{{Robust allocation results of the virtual inertia and damping parameters for IBRs.}}
		\label{fig.case2}
	\end{figure}
    \begin{figure}[h]
		\centering
		\includegraphics[width=0.7\linewidth]{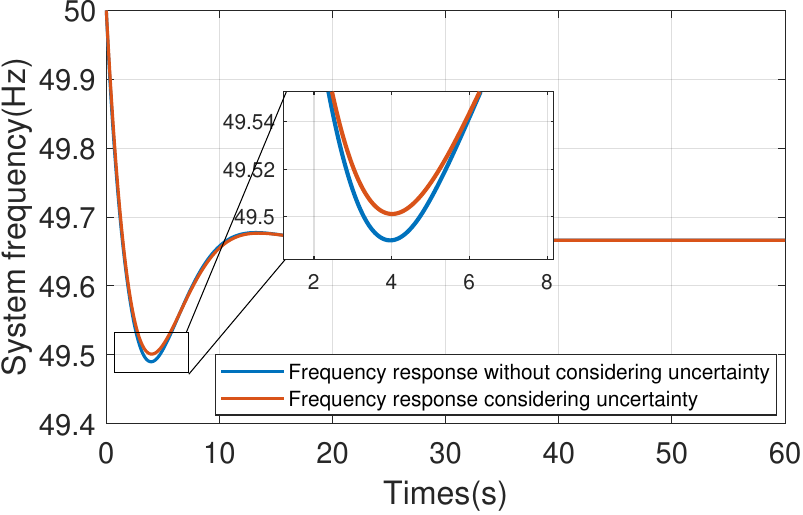}
		\caption{{Frequency responses under scheduling schemes with and without uncertainty.}}
		\label{fig.case22}
	\end{figure}
    
	\subsection{Case II: Robust Reserve Allocation Considering Uncertainty}
    
    In this section, the risk that the VPP cannot provide enough reserve of frequency response is considered. The fluctuation range set for each IBR adjustable capacity ($P_i^\text{rated}$) is [±20\%,±10\%,±9\%,±4\%,±5\%,±30\%,±5\%,±5\%]. Based on this, the VPP considers the potential maximum fluctuation in IBR power and reschedules the virtual inertia/damping parameters (shown in Fig.~\ref{fig.case2}).

    Under an actual fluctuation [-16.7\%,-9.1\%,0,0,0,-25\%,0,0], for example, the rescheduling results satisfy the system's frequency regulation requirements, and the VPP's economic performance remains unaffected by punishment. In contrast, the results of the deterministic model (shown in Fig.~\ref{fig.case1}) fail to fully meet the frequency regulation requirements (shown in Fig.~\ref{fig.case22}), resulting in a 2.54\% loss in VPP economic performance due to the punishment defined in (\ref{financial_P}).
    
    \subsection{Sensitive Analysis}
	Further parameter sensitivity analysis is conducted to study the impact of $H_{0}$ and $\Delta P$ on the cumulative energy of the proposed ResMin strategy to the ResPeak strategy without considering uncertainty. The parameter $H_{0}$ refers to the inertia of the power grid provided by SGs. Besides, the parameter $\Delta P$ refers to the amplitude of the power disturbance. For analysis of the grid's inertia, the varying range is set as $H_{0}=2, 3, 4, 5, 6, 7, 8, 9 $ (s) and the disturbance $\Delta P$ is set as $0.25$ (p.u.). While for analysis of power disturbance, the varying range is set as $\Delta P=0.21, 0.23, 0.25, 0.27, 0.29, 0.31, 0.33, 0.35 $ (p.u.) and the inertia is set as $5$ (s).
	
\begin{figure}[h]
		\centering
		\includegraphics[width=0.8\linewidth]{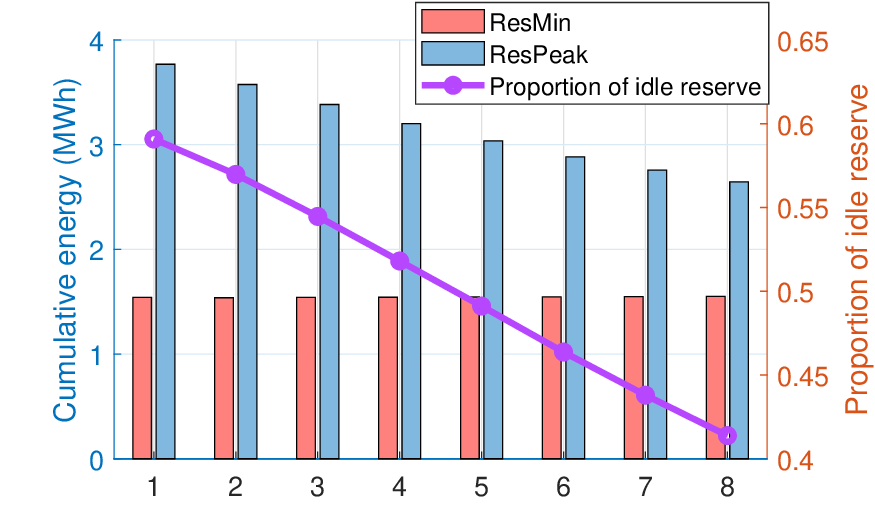}
		\caption{The impact of inertia parameter of the grid ($H_{0}$) on the economic benefit increment (from left to right are $H_{0}=2, 3, 4, 5, 6, 7, 8, 9 (s)$).}
		\label{fig.case3}
	\end{figure}

\begin{figure}[h]
		\centering
		\includegraphics[width=0.8\linewidth]{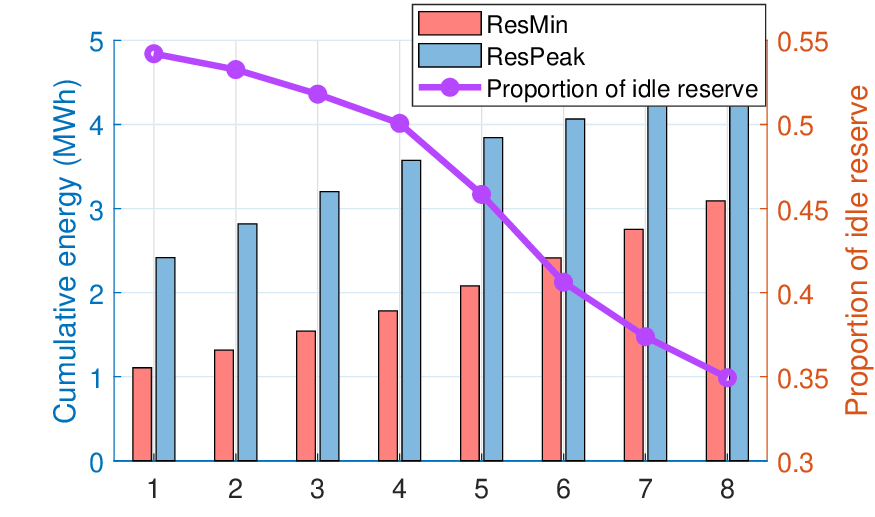}
		\caption{The impact of the power imbalance ($\Delta P$) on the economic benefit increment (from left to right are $\Delta P_{e}=0.21, 0.23, 0.25, 0.27, 0.29, 0.31, 0.33, 0.35 (p.u.)$).}
		\label{fig.case4}
	\end{figure}

	As shown in Fig. \ref{fig.case3}, a grid with lower inertia means that the VPP needs to provide more virtual inertia~($H_\text{VPP}$) under the same power disturbance. The larger $H_\text{VPP}$ has significant impacts on the peak values of the active power injections of the VPP, which leads to the prominent overshoot characteristic and corresponding proportion of idle reserve. Additionally, larger power disturbances necessitate increased power injections from the VPP, consistent with both the ResMin and ResPeak strategies. However, as illustrated in Fig. \ref{fig.case4}, the cumulative energy demand from ResMin increases at a faster rate compared to ResPeak. Consequently, the proportion of idle reserves decreases as the power imbalance $\Delta P_{e}$ increases.

	\section{Conclusion} \label{sec:conclusion}
	
	This paper introduces an economic reserve strategy aimed at minimizing the grid-side frequency regulation requirements for the VPP and enhancing its real-time electricity market benefits. An eighth-order state-space model is developed to describe the dynamics of the VPP interaction system, deriving the decay rate constraint for response speed. Additionally, a reduced-order model is proposed to map the relationship between the active power disturbance and the frequency response, identifying three safety constraints. Based on these models, the feasible region of the VPP's regulation reserve is defined, and a sequential algorithm is proposed to determine the minimal reserve. The optimal allocation of the VPP-level reserve is achieved through an energy-based optimization model, which is convex and capable of finding the optimum solution.
	
	The results of case studies suggest that:

	\begin{enumerate}
		\item {Compared with the reserve decision strategy based on peak value, the proposed method shows a significant improvement in financial gains by releasing idle reserves over 51.88\%.}
		
		\item {The proposed energy-based reserve allocation achieves the optimal gains for the VPP, which increases the even allocation and the proportional allocation by 4.91\% and 5.43\% economically.}
		
		\item {Certain fluctuations in the regulation power capacity of IBRs may hinder the VPP from fully meeting the regulation requirements, resulting in a 2.54\% loss in economic performance. This issue can be addressed by designing a robust rescheduling scheme appropriately.}
	\end{enumerate}

    Future research will explore reserve market schemes by introducing VPP gain bonuses for maintaining idle reserves. Additionally, it will focus on active strategies to systematically address scheduling under uncertainties arising from disturbances and renewable energy generation.
	
	\section{Appendix}
	
	\subsection{The Matrices of the Eighth-Order Model}
	The state matrix $\bf{A}$ and input matrix $\bf{Bu}$ are shown in (\ref{eq.ss-A}) and (\ref{eq.ss-B}) respectively.
	
	\begin{equation}
		{\bf{A}} = \left( {\begin{array}{*{20}{c}}
				{{{\bf{A}}_{\bf{1}}}}&{{{\bf{A}}_{\bf{2}}}}\\
				{{{\bf{A}}_{\bf{3}}}}&{{{\bf{A}}_{\bf{4}}}}
		\end{array}} \right)
		\label{eq.ss-A}
	\end{equation}
	
	\begin{equation}
		{{\bf{A}}_{\bf{1}}} = \left( {\begin{array}{*{20}{c}}
				{ - K_p^PM}&{K_p^I}&{K_p^PM}&0\\
				{ - M}&0&M&0\\
				0&0&{ - \frac{{{D_0}}}{{2{H_0}}}}&0\\
				{K_r^I(2{H_{{\rm{VPP}}}}K_p^PM - {D_{{\rm{VPP}}}})}&{ - 2K_r^I{H_{{\rm{VPP}}}}K_p^I}&{ - 2K_r^I{H_{{\rm{VPP}}}}K_p^PM}&0
		\end{array}} \right)
	\end{equation}
	
	\begin{equation}
		{{\bf{A}}_{\bf{2}}} = \left( {\begin{array}{*{20}{c}}
				0&0&0&0\\
				0&0&0&0\\
				0&{\frac{1}{{2{H_0}}}}&0&{ - \frac{1}{{2{H_0}}}}\\
				0&{ - K_r^I}&0&0
		\end{array}} \right)
	\end{equation}
	
	\begin{equation}
		{{\bf{A}}_{\bf{3}}} = \left( {\begin{array}{*{20}{c}}
				0&0&0&0\\
				{\frac{{K_r^P(2K_p^P{H_{{\rm{VPP}}}}M - {D_{{\rm{VPP}}}})}}{{{L_1}}}}&{ - \frac{{2(K_r^PK_p^I{H_{{\rm{VPP}}}})}}{{{L_1}}}}&{ - \frac{{2(K_r^PK_p^P{H_{{\rm{VPP}}}}M)}}{{{L_1}}}}&{\frac{1}{{{L_1}}}}\\
				0&0&0&0\\
				0&0&{\frac{R}{{{T^{{\rm{SG}}}}}}}&0
		\end{array}} \right)
	\end{equation}

	\begin{equation}
		{{\bf{A}}_{\bf{4}}} = \left( {\begin{array}{*{20}{c}}
				0&0&{ - K_r^I}&0\\
				0&{ - (\frac{1}{{{L_1}}}K_r^P + \frac{1}{{{L_1}}}{R_1})}&0&0\\
				{\frac{1}{{{L_1}}}}&0&{ - \frac{1}{{{L_1}}}(K_r^P + {R_1})}&0\\
				0&0&0&{ - \frac{1}{{{T^{{\rm{SG}}}}}}}
		\end{array}} \right)
	\end{equation}
	
	\begin{equation}
		\setlength{\abovedisplayskip}{3pt}
		\setlength{\belowdisplayskip}{3pt}
		{\bf{Bu}} = \left( {\begin{array}{*{20}{c}}
				0\\
				0\\
				{\frac{{P^{G} - P^{L}+D_{0}}}{{2{H_0}}}}\\
				{K_{r}^I(i_{0}^V + {D_\text{VPP}}{f_0}-{D_\text{VPP}}{f_\text{DB1}})}\\
				0\\
				{\frac{1}{{{L_1}}}K_{r}^P(i_{0}^V + {D_\text{VPP}}{f _0} - {D_\text{VPP}}{f_\text{DB1}})}\\
				0\\
				{\frac{R}{T^\text{SG}}({f_\text{DB2}} - f_{0})}
		\end{array}} \right)
		\label{eq.ss-B}
	\end{equation}

	\subsection{Time-Domain Expression of Frequency Response}
	To derive the time-domain expression of system frequency response, the differential equations (\ref{td-SG}), (\ref{td-VPP}) and (\ref{swingnew}) are transferred into s-domain in (\ref{appendix-1}) - (\ref{appendix-3}) and solved by Inverse Laplace transform.
	\begin{equation}
		\setlength{\abovedisplayskip}{3pt}
		\setlength{\belowdisplayskip}{3pt}
		\left\{ \begin{aligned}
			& \Delta f=\left( \Delta {{P}_\text{dis}}(s)+\Delta {{P}_{m}}(s) \right)\frac{1}{2{{H}_{0}}s+{{D}_{0}}} \\ 
			& \Delta P_{m} = -2{{H}_\text{VPP}}s\cdot \Delta f(s),0\le t \le t_{DB1}\\
			& \Delta {{P}_{m}}=-\left( \frac{R}{1+{{T}^\text{SG}}s}+2{{H}_\text{VPP}}s+{{D}_\text{VPP}} \right)\Delta f(s)\\
			&-\frac{R}{1+{{T}^\text{SG}}s}\cdot \frac{{{f}_\text{DB1}}}{s}-{{D}_\text{VPP}}\cdot \frac{{{f}_\text{DB2}}}{s}, t \ge t_{DB2} \\ 
		\end{aligned} \right.
		\label{appendix-1}
	\end{equation}

	For $0\le t \le t_{DB1}$, the system frequency is formulated as in (\ref{appendix-2}).
	\begin{equation}
		\setlength{\abovedisplayskip}{3pt}
		\setlength{\belowdisplayskip}{3pt}
		\Delta f(s) =  \frac{\Delta P_\text{dis}(s)}{2(H_{0}+H_\text{VPP})s+D_{0}}
		\label{appendix-2}
	\end{equation}
	
	For $t \ge t_{DB2}$, the system frequency is formulated as in (\ref{appendix-3}).
	\begin{equation}
		\setlength{\abovedisplayskip}{3pt}
		\setlength{\belowdisplayskip}{3pt}
		\Delta f(s) =  \frac{{(1 + {T^\text{SG}}s)(\Delta P_\text{dis}(s) - {D_\text{VPP}}{f_\text{DB2}}) - R \cdot {f_\text{DB1}}}}{{s(R + (2Hs + D)(1 + {T^\text{SG}}s))}}
		\label{appendix-3}
	\end{equation}

    \section*{Acknowledgments}
    This work was supported by the National Key Research and Development Program of China under Grant No. 2024YFB4207200 and the National Natural Science Foundation of China under Grant No. U24B2085.
    
	\bibliographystyle{elsarticle-num-names} 
	\bibliography{cas-refs}

\end{document}